\documentclass[fleqn,12pt,twoside]{article}
\usepackage{espcrc1}

\usepackage{graphicx}
\usepackage[figuresright]{rotating}

\newcommand{\AmS}{{\protect\the\textfont2
  A\kern-.1667em\lower.5ex\hbox{M}\kern-.125emS}}
\newcommand{\mvn}{${\cal M}_{VN}$ }
\newcommand{\ro}{$\rho$ }
\newcommand{\om}{$\omega$ }
\newcommand{\plab}{${\bf p}$ }

\title{Vector Mesons and Baryon Resonances in Nuclear Matter}

\author{M. Post\address[MCSD]{Institut f\"ur Theoretische Physik, 
        Heinrich-Buff Ring 16, \\ 
        35392 Giessen, Germany}%
        \thanks{Work supported by BMBF}
        and 
        U. Mosel \addressmark }

\begin{document}

\maketitle

\begin{abstract}
We calculate the effect of many-body interactions in nuclear matter on the
spectral function of $\rho$ and $\omega$ meson. In particular, we focus 
on the role played by baryon resonances in this context.
\end{abstract}

\section{Introduction}
Recently the dilepton spectra from the CERES collaboration \cite{ceres95},
which indicate a mass shift of $\rho$ and $\omega$ mesons, have 
been interpreted as a signature for the formation of a new
state of matter, the Quark Gluon Plasma. 
Within this picture, the medium modifications are a consequence of 
the restauration of chiral symmetry. For example, 
in the Brown-Rho scaling scenario \cite{br} the mass of the vector mesons scales 
with the
chiral condensate which is supposed to drop at finite density and 
temperature. However, in a dense and hot medium a variety of conventional many-body 
phenomena with considerable impact on the in-medium properties of the 
vector mesons may occur \cite{klingl,rw,fp,pp98,plm,pm,frilu}. 
In this contribution we present a calculation of these effects 
at finite density and at $T = 0$. In particular, we 
concentrate on the effects played by the excitation of baryon resonances, which
in turn are modified themselves in the nuclear environment. The interplay between
resonances and vector mesons is accounted for in a selfconsistent approach.

\section{The Model for VN Scattering}
In the following we will focus on the calculation of the in-medium 
spectral function $A$ of both \ro and \om meson. 
The spectral function is 
defined as the imaginary part of the propagator and can be directly 
interpreted as the mass distribution of a particle. Whereas in the vacuum 
the spectral functions are controlled by the well-known pionic decays of the 
vector mesons, new decay channels open up in the 
nuclear medium due to the possibility of scattering with surrounding 
nucleons. This is formally expressed in the low-density 
approximation, which relates 
the in-medium selfenergy to the vacuum forward scattering 
amplitude \mvn on the nucleon. Experimental constraints on these amplitudes
exist only through an analysis of vector meson production. 
In the case of the 
\ro meson, a further complication arises from its large decay width, 
which makes the experimental identification difficult. 
Therefore, in principle \mvn has to follow from a coupled-channel 
analysis of $\pi\,N$ (and $\gamma\,N$) scattering. 
This is a formidable task and one might
wonder if the gross information on \mvn can not be obtained from a
more elementary model. 
The results of such an analysis from Manley {\it{et al}} \cite{man} indicate 
that resonances dominate $\pi\,N$ scattering. We therefore
approximate \mvn within a resonance model for both \ro and \om meson.
Experimental information then enters directly through the coupling constant 
$f_{RNV}$ at the resonance vertex. 

Let us point out the implications
of resonance scattering on the spectral function.
Through the excitation of a resonance the vector meson
can convert into a resonance-hole state in the same manner as is well known
from the pion Delta-hole model. 
The in-medium spectral function therefore contains additional peaks
corresponding to these new states. From basic kinematical
considerations it follows that their position moves down to lower invariant masses
with increasing momentum \plab of the vector meson relative to nuclear matter, thus
producing a momentum dependent spectral function. Note also, that in nuclear matter
transversely and longitudinally polarized vector mesons are modified differently
for $\plab  \ne 0$.

\section{Results for the \ro Meson}

We turn now to the results for the \ro meson. 
We consider all resonances which have been assigned a coupling
to this channel in the analysis of Manley {\it et al} \cite{man}. As it turns out,
by far the strongest coupling constants are obtained for some subthreshold
resonances with masses $m_R < m_N + m_{\rho} \approx 1.7$ GeV \cite{pp98,plm}. 
The $N\,\rho$ decay of these 
resonances is phase-space suppresed and can only proceed via the low-mass tail
of the \ro spectrum. Clearly, to accomodate the extracted
decay widths, large coupling constants are mandatory. For the same reason we find
the contribution from high lying resonanes in general to be negligible. 
\begin{figure}[h]
\includegraphics[width=8cm]{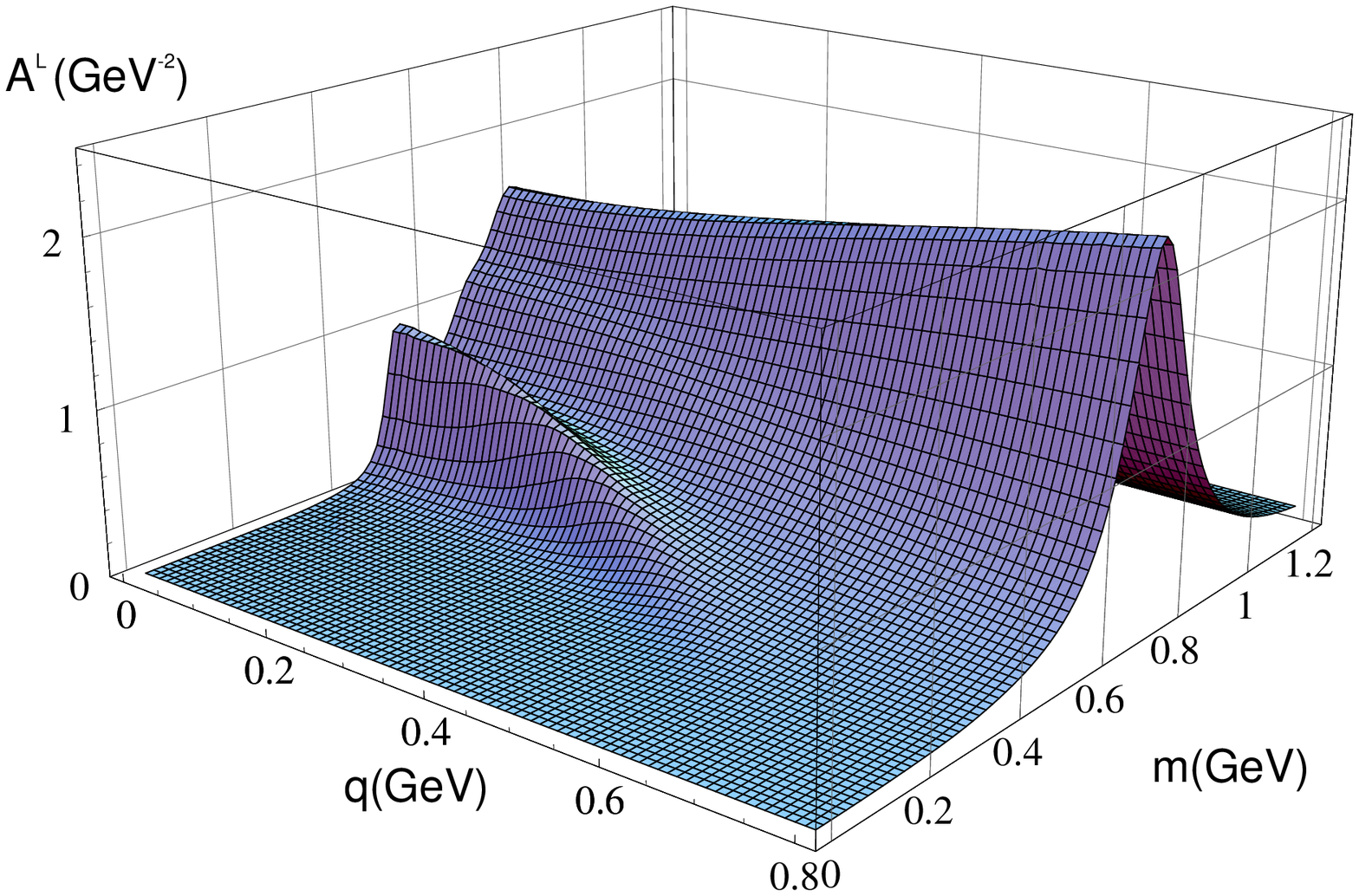}
\includegraphics[width=8cm]{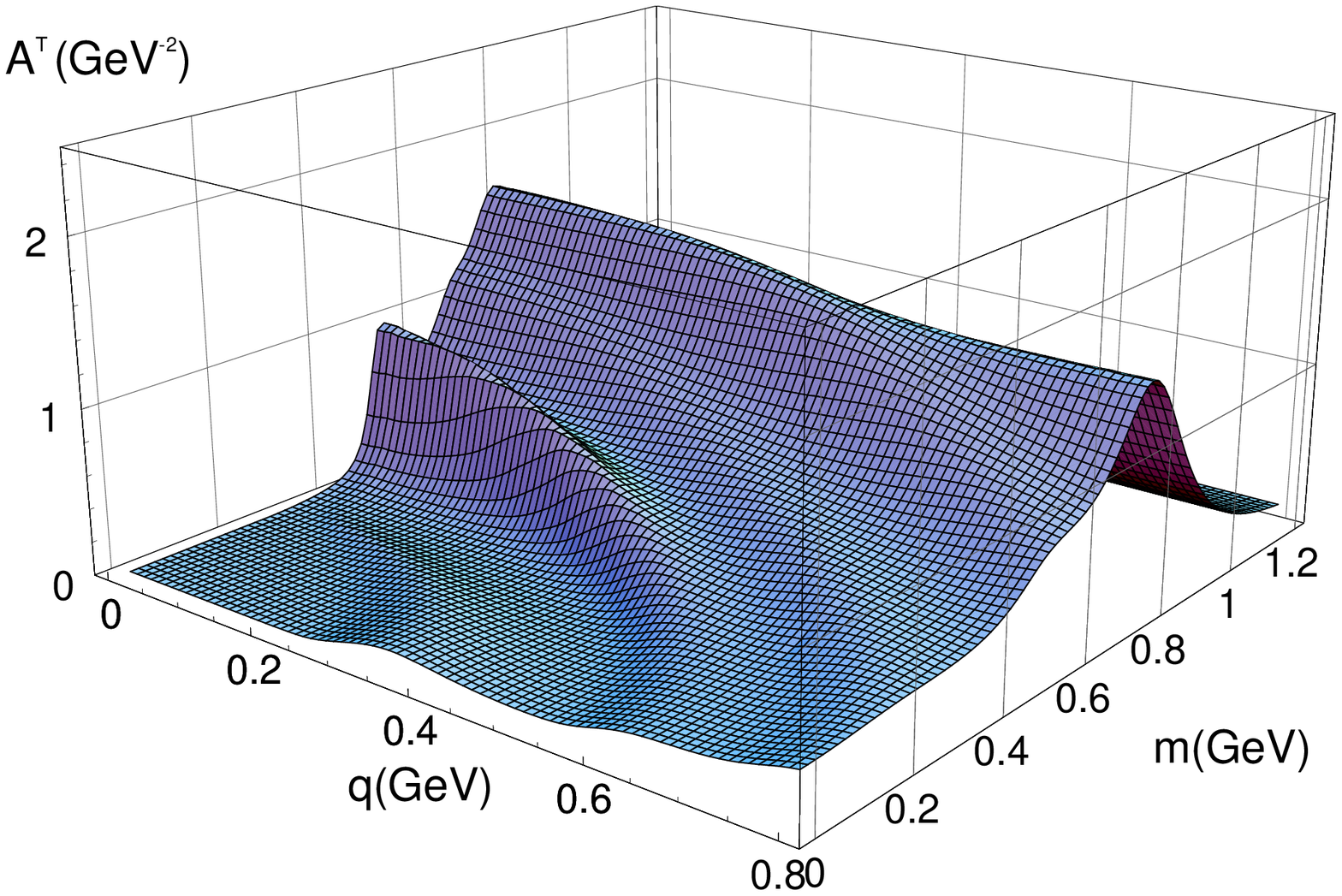}
\caption{
\label{rhospec} Left: Longitudinal spectral function of the \ro in nuclear matter
as a function of its invariant mass $m$ and \plab. Right: same for the transverse
spectral function.}           
\end{figure}
\vspace{-0.5cm}
This means for the spectral function that a substantial amount of strength
is relocated to lower invariant masses, 
as is shown in figure \ref{rhospec}. At low momenta
both the transverse and the longitudinal spectral function are dominated from
the $N^{\star}(1520)$ resonance which has a $N\,\rho$ decay width of about $25$ MeV
and, therefore, a large coupling constant $f_{RN\rho}$.
At higher momenta, transverse and longitudinal \ro mesons behave differently:
in the longitudinal channel the free spectral function is nearly recovered, whereas
the transverse channel exhibits a strong broadening.

\section{Baryon Resonances in Nuclear Matter}
\begin{figure}[h]
\includegraphics[width=16cm]{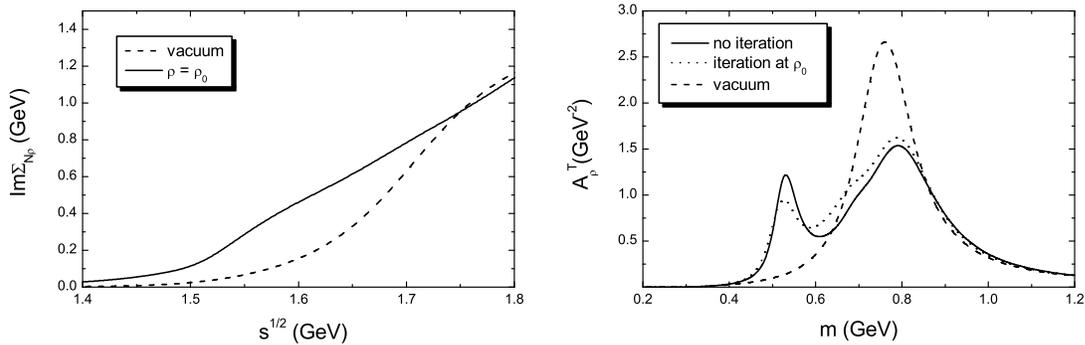}
\vspace{-2cm}
\caption{
\label{resmed} Left: In-medium decay width of the $N^{\star}(1520)$ at
density $\rho = \rho_0$ and $\plab=0$. 
Right: Effect of the selfconsistent calculation on the
spectral function of the $\rho$ ($\plab=0$).}
\end{figure}

As explained in the last section, the coupling of \ro mesons to resonance-hole
states leads to a substantial shift of spectral strength 
down to lower invariant masses. We also pointed out that the vacuum decay 
of the subtrheshold
resonances is phase space suppressed. One might therefore expect a large
feedback effect of a lighter \ro meson on the in-medium properties 
of baryon resonances \cite{pp98}.
We calculate this response in a self-consistent fashion by calculating 
the $N\,\rho$ decay width
of the resonances with the in-medium spectral function of the \ro.
In a different language, this amounts to a calculation of the collisional 
broadening of the resonance through the exchange of a medium modified \ro meson.
On top of the collisional broadening we also include, of course, Pauli-blocking
for all decay channels.
As shown in figure \ref{resmed} for the $N^{\star}(1520)$, the in-medium decay width
is strongly enhanced, at the pole mass we find $\Gamma_{N\,\rho} \approx 100$ MeV
for a $N^{\star}(1520)$ at rest. At larger momenta the decay width increases as the
effects from Pauli blocking get less important. This broadening leads
to a reduction of the influence of the resonances on the \ro spectral function.
\vspace{-1cm}

\vspace{-1cm}
\section{Results for the \om Meson}
\begin{figure}[t]
\includegraphics[width=16cm]{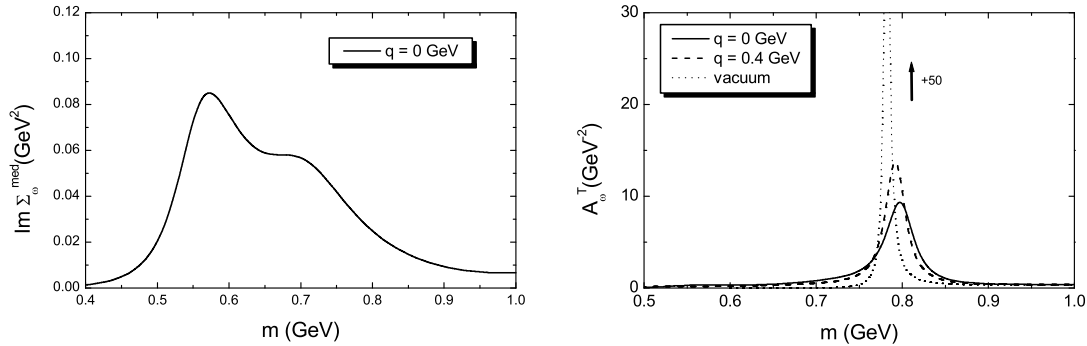}
\vspace{-1.5cm}
\caption{
\label{om}{Left: Spectral function of the \om meson at $\rho = \rho_0$ at rest 
and with momentum 0.4 GeV relative to nuclear matter. Right: Im $\Sigma$ for
a \om meson at rest.}\vspace{-1cm}}
\end{figure}

A coupling of baryon resonances to the $N\,\omega$ channel has so far not been
experimentally established. However, the previous discussion of the in-medium \ro meson 
suggests that resonances have a strong impact on the 
in-medium \om meson as well.
In order to obtain an estimate for the coupling constants $f_{RN\omega}$, 
we perform a VMD analysis of the   
electromagnetic (EM) decay vertex of the resonances, using the helictiy 
amplitudes of Arndt \cite{arndt}. By 
decomposing the EM coupling into an isoscalar and an isovector part
both $f_{RN\rho}$ and $f_{RN\omega}$ can be accessed. 
In the isovector channel we compare the VMD results 
with our fits to the hadronic $N\,\rho$ width and find reasonable agreement.
For the \om the VMD analysis predicts a sizeable coupling of the $N^{\star}(1535)$,
$N^{\star}(1650)$ and $N^{\star}(1520)$ resonances. Within in a resonance model we 
confirmed that the obtained couplings are compatible with experimental data
on $\pi(\gamma)\,N \rightarrow \omega\,N$ \cite{pm}. 
However, the couplings constants are smaller than in the
case of the \ro. This leads to less pronounced resonance peaks in the 
spectral function, see figure \ref{om}. We find that 
the in-medium width of the \om is about $40$ MeV and that its mass goes slightly up
by about 20 MeV. 

\section{Summary $\&$ Outlook}
We have presented a calculation of the spectral function of \ro and \om 
mesons at finite density and $T=0$ within a resonance model. We also considered
the feedback effect on the baryon resonances caused by the change in the spectral
distribution of the \ro. However, here only the collisional broadening of the 
resonance, but not the corresponding mass shift was calculated. We are currently
working on a more complete calculation of these effects.

\end{document}